\documentclass[journal,transmag]{IEEEtran}
\usepackage[cmex10]{amsmath}
\usepackage[pdftex]{graphicx}
\usepackage[caption=false,font=footnotesize]{subfig}

\begin{document}

\title{Effects of Elastic Dephasing on Scaling of ultra-small Magnetic Tunnel Junctions}

\author{\IEEEauthorblockN{Debasis Das\IEEEauthorrefmark{1},
Ashwin Tulapurkar\IEEEauthorrefmark{1},
Bhaskaran Muralidharan\IEEEauthorrefmark{1}, 
~\IEEEmembership{Member,~IEEE}}
\IEEEauthorblockA{\IEEEauthorrefmark{1}Department of Electrical Engineering, Indian Institute of Technology Bombay, Mumbai 400076, India}

\thanks{Corresponding author Debasis Das (e-mail: ddas@ee.iitb.ac.in).}}

\IEEEtitleabstractindextext{%
\begin{abstract}
The study of the effects of scaling on magnetic tunnel junction (MTJ) devices has become an important topic in the field of spin-based memory devices. Here, we investigate the effect of elastic dephasing on trilayer and pentalayer MTJ considered at small transverse cross-sectional areas using the non-equilibrium Green's function spin transport formalism. We consider the structures with and without dephasing effects and clearly point out as to how the tunnel magnetoresistance effect gets affected by dephasing. We attribute the trends noted by analyzing the transmission spectra and hence the currents across the devices. Although dephasing affects the TMR values for both devices, we note that the obtained TMR values are still in a reasonable range that may not hinder their usability for practical applications.
\end{abstract}

}

\maketitle

\IEEEdisplaynontitleabstractindextext

\IEEEpeerreviewmaketitle

\section{Introduction}

\IEEEPARstart{S}{pintronic} devices featuring magnetic tunnel junctions (MTJs) have attracted a lot of attention due to low power operation, non-volatility, and the possibility of writing data using spin currents via the spin-transfer torque (STT) effect. The principal technological applications are spin based magnetic random access memories (MRAMs)\cite{huai2008spin}, spin torque nano-oscillators\cite{Houssameddine2007,Sharma2015,Kiselev2003}, and spin-based logic devices\cite{Nikonov2011}. The basic structure of the STT-MRAM cell is the MTJ which consists of two ferromagnetic (FM) layers separated by a thin oxide layer. The magnetization of one FM layer called the reference layer (RL) is kept fixed via an antiferromagnetic exchange coupling, whereas the magnetization of the other FM layer called the free layer (FL) can be rotated by some externally applied magnetic field or via current-induced STT. Reading information from the MTJ depends on the relative orientation between the FL and RL, and the quality of an MTJ's electrical read-out/write process is determined by the tunnel magnetoresistance ratio (TMR) defined as $TMR=(R_{AP}-R_P)/R_P$, where $R_{AP}$ and $R_P$ are the resistance of the device when magnetizations of the FL and RL are in the antiparallel configuration (APC) or the parallel configuration (PC), respectively.

While several groups have reported on how to enhance the TMR \cite{parkin2004giant,wang2010coherent} in typical trilayer devices, in recent years, pentalayer devices which feature resonant tunneling have also been proposed \cite{chatterji2014enhancement,sharma2016ultrasensitive}. The resonant tunneling MTJs (RTMTJs) offer an ultrahigh TMR due to the resonant spin filtering phenomenon \cite{sharma2016ultrasensitive}. As the scaling of the devices has become an important field of study, here we try to explore as to how the device characteristics of MTJs and RTMTJs are affected at an ultrasmall scale. From literature, it is well known that magnets with out-of-plane anisotropy switch with a lower critical current than those with in-plane anisotropy \cite{mangin2006current,meng2006spin}. Hence, for all our calculations, we choose devices featuring magnets with out-of-plane anisotropy. Although in-plane anisotropy can also be assumed as long as the switching of FL is not considered, due to the fact that, both in-plane and out-of-plane magnetization gives rise to the same TMR.

In our recent work \cite{das2018scaling} we have shown how scaling affects the device characteristics of MTJ and RTMTJ structures by varying the cross-sectional area along the transverse direction, from 25 $nm^2$ to 10000 $nm^2$ using spin-resolved nonequilibrium Green's function (NEGF) approach. We have noted that the highest TMR was achieved at the smallest cross-sectional area, i.e., at 25 $nm^2$, while maintaining a low resistance-area (RA) product. As the quality of an MTJ device depends on the TMR value, we can conclude that for a better MTJ, it is required to fabricate it with the smallest possible transverse cross-sectional area. So to see the effects of dephasing on MTJ as well as RTMTJ structures, we select the device with the smallest cross-sectional area. In practice, there are several sources which degrade the device performance, and these can be modeled qualitatively by introducing dephasing \cite{datta2} within the device region. There are basically two types of dephasing \cite{golizadeh2007nonequilibrium}. One which destroys the phase but conserves the momentum, where electron-electron interaction is the source of this kind of dephasing. On the other side interface roughness, impurity scattering, acoustic phonon scattering introduces elastic dephasing which destroys both phase and momentum. In this work, the effect of elastic scattering on MTJ and RTMTJ is analyzed by varying strength of dephasing.

\section{Device Modeling}
\subsection{Device Structure}
The device structures used for this simulation are shown in Fig. 1. In the case of the MTJ, a thin MgO layer is sandwiched between the FL and RL, whereas in the RTMTJ a heterostructure consisting MgO-Semiconductor(SC)-MgO is sandwiched between the FL and RL. In our simulations, we have assumed that the FM layers to have perpendicular anisotropy, although in-plane anisotropy produces same TMR, as it depends on the relative angle between the magnetization of FL and RL. The magnetization of the RL is fixed along the y-direction whereas the FL is assumed to be confined in the y-z plane but making an angle $\theta$ with the y-direction as a result of thermal noise. For more accurate results, we performed a basis transformation\cite{yanik2007quantum} as the magnetization of both FM layers are not collinear. For all the devices, the direction of the electronic transport is assumed to be along the y-direction. 
\begin{figure}[h]
    \subfloat[]{\includegraphics[scale=0.22]{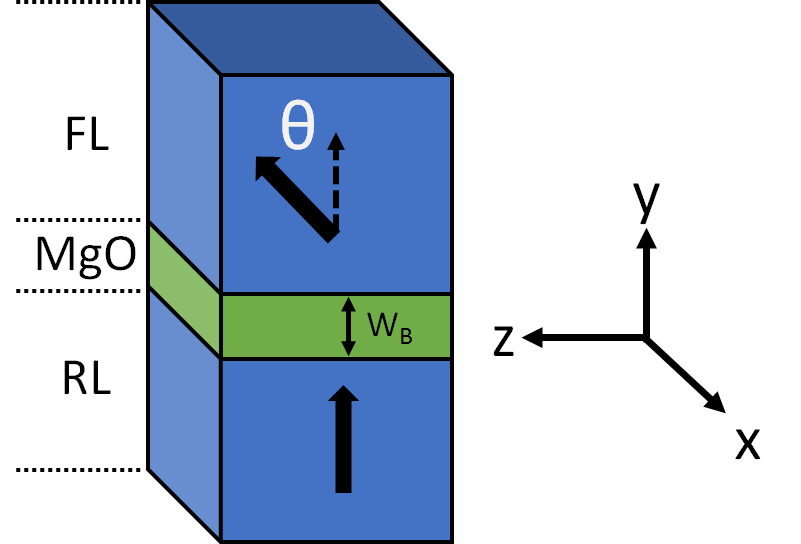}}
    \hfill
    \subfloat[]{\includegraphics[scale=0.18]{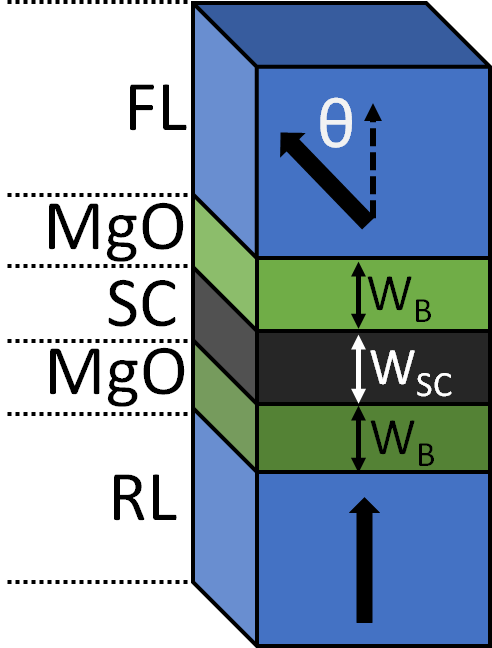}}
    \caption{(a) Schematic of a trilayer MTJ device. An MgO layer is sandwiched between the FL and RL. $W_B$ is the width of the MgO layer. (b) Schematic of the pentalayer RTMTJ. Between the two FM layers, a heterostructure of MgO-Semiconductor(SC)-MgO is sandwiched. $W_{SC}$ is the width of the semiconductor layer. }
    \label{Device_structure}
\end{figure}

\subsection{Simulation Method}\label{method}
We have assumed ballistic transport in all the devices due to the small length scales involved (details are in \textit{device parameter} section) along the transport direction which justifies the uncoupled mode space approach assumed for this simulation work. To simulate, using mode space approach, each transverse mode is calculated by solving the 2D Schr\"{o}dinger equation along the transverse cross-section assuming that the wave vector \textbf{k} for each transverse mode is conserved. For calculating the total charge current flowing through the device, first we solve the 1-D NEGF equations for each transverse mode and then sum it over all the transverse modes. The NEGF calculation starts with the energy-resolved Green's function matrix [G(E)] which is obtained from device Hamiltonian matrix and it is given by
\begin{equation}
\left[ G(E)\right] =\left[ EI-H_0-U-\Sigma_T-\Sigma_B\right]^{-1}\label{G_eq}
\end{equation}
where, $[H_0]$ is the Hamiltonian matrix calculated from an effective mass tight binding approach, $[U]$ is the Coulomb charging matrix,  and $[I]$ is the unitary matrix whose size depends on the number of lattice points along the transport direction. Here, $\left[ \varSigma_T\right]$ and $\left[ \varSigma_B\right]$ are the self-energy matrices of the top (FL) and bottom (RL) layers respectively.

\begin{figure}[h]
    \begin{center}
        \subfloat[]{\includegraphics[scale=0.092]{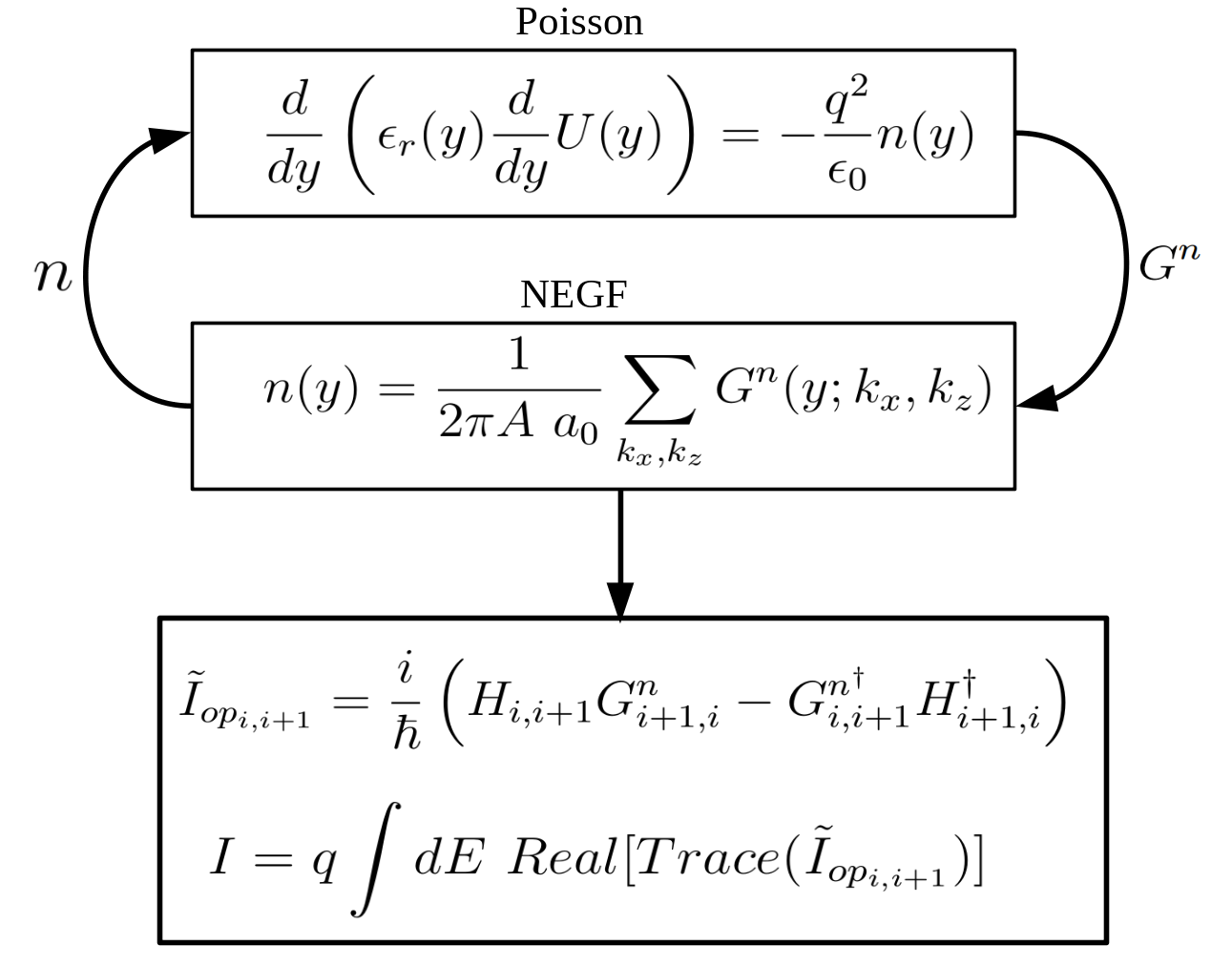}}
        \hfill
        \subfloat[]{\includegraphics[scale=0.16]{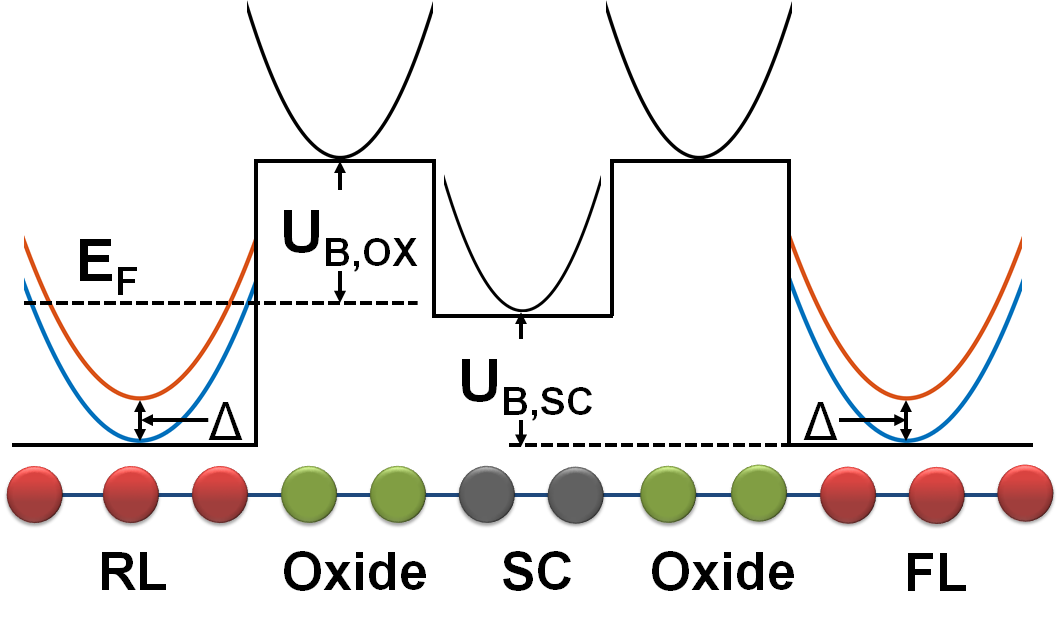}}
    \end{center}
    \caption{(a) Simulation engine for the non-equilibrium Green's function formalism self consistently coupled with Poisson's equation to capture the RTMTJ potential profile. (b) Schematic of the energy band diagram of a pentalayer RTMTJ for a single transverse mode. The exchange splitting is $\Delta$ for the FM contact, $U_{B,OX}$ is the barrier height of the oxide layer above Fermi energy $E_F$, and $U_{B,SC}$ is the energy difference between the bottom of the conduction band of the semiconductor and the FM. Parabola with red line in RL/FL denotes the energy band for the up spin electron whereas, the blue one represents the down spin energy band. Parabola with black line in the oxide and semiconductor region denotes the overlap of the up and down spin energy bands. Chain of circular dots in the schematic diagram denote the atoms of one dimensional model assumed for any single transverse energy mode.}
    \label{engine_band}
\end{figure}

For a trilayer MTJ, an oxide layer is sandwiched between two FM layers and hence it can be assumed that there will be a linear potential drop in the oxide layer. However, due to the presence of the heterostructure (MgO-Semiconductor-MgO) in the RTMTJ case, we capture the potential drop inside the device accurately via the self-consistent NEGF-Poisson solver \cite{das2018scaling}. The potential can be obtained by solving the following equations self-consistently.  
\begin{align}
\frac{d}{dy}\left(\epsilon_r(y)\frac{d}{dy}U(y)\right)&=-\frac{q^2}{\epsilon_0}n(y)\label{U_eq}\\
n(y)&=\frac{1}{2\pi A\ a_0} \sum_{k_x,k_z} G^n(y;k_x,k_z), \label{n_eq}
\end{align}
where $a_0$ is the interatomic spacing, q is the electronic charge, $\epsilon_r$ and $\epsilon_0$ are relative and free space permittivity. $A$ is the transverse cross-sectional area, $n(y)$ is the electron density and $G^n(y;k_x,k_y)$ is the diagonal element of the electron correlation matrix given by
\begin{equation}
[G^n]=\int dE\ \left[ [A_T(E)]f_T(E)+[A_B(E)]f_B(E)\right].
\end{equation}
Here, $[A_{T,B}(E)]=G(E)\Gamma_{T,B}(E)G^{\dagger}(E)$ is the spectral function, $\Gamma_{T,B}(E)=i([ \Sigma_{T,B}(E)] - [ \Sigma_{T,B}^{\dagger}(E)])$ represents the spin-dependent broadening matrices of the top (T) and the bottom (B) contact respectively, and $f_T(E)$ and $f_B(E)$ represent the Fermi-Dirac distributions of the top and bottom contacts respectively. After solving $[U]$ self-consistently, the charge current operator $\tilde{I}_{op}$ between two neighboring lattice points $j$ and $j+1$ is calculated using
\begin{equation}\label{Iop_eq}
\tilde{I}_{op_{j,j+1}}=\frac{i}{\hbar}\left( H_{j,j+1}G^n_{j+1,j}-G^{n^{\dagger}}_{j,j+1}H^{\dagger}_{j+1,j}\right).
\end{equation}
where $H$ and $G^n$ are 2 $\times$ 2 matrices in spin space and hence $\tilde{I}_{op}$ is also a 2 $\times$ 2 matrix in spin space. The charge current calculated from the current operator is given by
\begin{equation}
I=q\int dE\ Real[Trace(\tilde{I}_{op_{j,j+1}})]
\end{equation}
The basic simulation method to solve self-consistent NEGF-Poisson equation and to deduce current is schematized in the Fig. \ref{engine_band} (a).
While considering elastic dephasing effects, the equations for the corresponding self-energy matrices and the in-scattering functions get modified as
\begin{align}
\left[\Sigma(E)\right]&=\left[\Sigma_T(E)\right]+\left[\Sigma_B (E)\right]+\left[\Sigma_0 (E)\right]\\
\left[\Sigma^{in} (E)\right]&=\left[\Gamma_T(E)\right]f_T(E)+\left[\Gamma_B(E)\right]f_B(E)+\left[\Sigma^{in}_0 (E)\right]
\end{align}
where $\Sigma_0(E)$ and $\Sigma^{in}_0(E)$ are given by 
\begin{align}
\Sigma_0(E)&= D\otimes [G(E)]\\
\Sigma^{in}_0(E)&=D\otimes [G^n(E)]
\end{align}
Here, [D] denotes the correlation function involving the dephasing potentials \cite{golizadeh2007nonequilibrium}. For elastic dephasing which destroys phase and momentum, [D] matrix is chosen such that $D(i,j)=D_0 \delta_{ij}$, where magnitude of $D_0$ denotes strength of dephasing.

\subsection{Device Parameters}
For the simulation, device parameters were chosen from previously published articles \cite{Datta2012,chatterji2014enhancement}. For both device structures, we use CoFeB as the FM material for both FL and RL with an exchange splitting $\Delta$=2.15 eV and Fermi energy $E_F$ at 2.25 eV \cite{Datta2012}. Parameter details are shown in Fig. \ref{engine_band} (b). The effective mass for the MgO is chosen to be $m_{OX}$=0.18$m_0$, whereas the barrier height from the Fermi level $U_{B,OX}$=0.76 eV is chosen \cite{Datta2012}. The width of each oxide layer in the MTJ, as well as the RTMTJ, is taken to be $W_B$=1 nm. The width of the SC in the RTMTJ is $W_{SC}$=1 nm, with the effective mass $m_{SC}$=0.38$m_0$, the well width $U_{B,SC}$=-0.45 eV \cite{sharma2016ultrasensitive} and $m_0$ is the free electron mass. For the TMR calculation, the voltage between two FM layers is kept at a very low value (10 mV), so that itinerant spins do not affect the magnetization of the free layer FM.


\section{Results and Discussion}
\subsection{Trilayer MTJ}
In our previous work\cite{das2018scaling}, we demonstrated how scaling affects TMR by varying the transverse cross-sectional area of the device from 25 $nm^2$ to 10000 $nm^2$.  The TMR is calculated as 
\begin{equation}\label{TMR_eq}
TMR=\frac{I_P-I_{AP}}{I_{AP}}=\frac{I_P}{I_{AP}}-1,
\end{equation}
where $I_P$ and $I_{AP}$ are the charge currents of the device in the PC and the APC respectively. In both structures, we have seen that the TMR remains constant at larger areas, but it shoots up rapidly if the area is reduced below a certain value. This result suggested that the highest TMR was obtained at 25 $nm^2$. As the main criteria to design the MTJs is to obtain high TMR, and it was obtained at the smallest area, so we limit all of our simulations to the smallest area, i.e., at 25 $nm^2$ to save computational cost. To explore the effects of dephasing, we varied the strength of dephasing $D_0$ \cite{sharma2018role,datta2}. 

\begin{figure}[h]
    \subfloat[]{\includegraphics[scale=0.28]{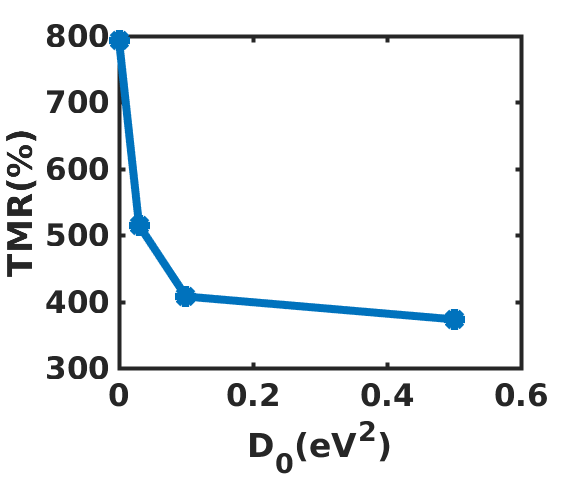}}
    \hfill
    \subfloat[]{\includegraphics[scale=0.28]{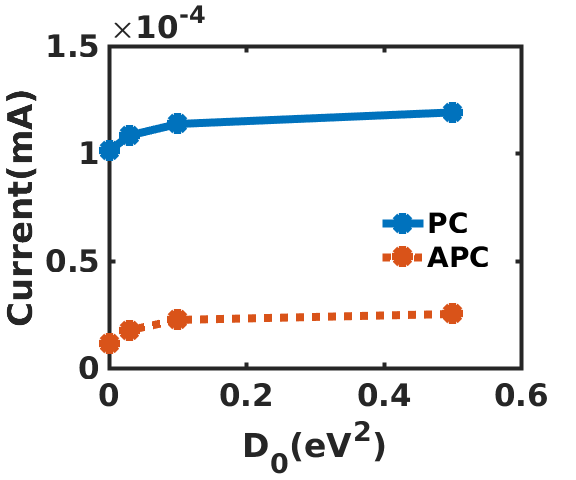}}
    \caption{Dephsing effects on trilayer MTJ structures. (a) Variation of TMR with the strength of dephasing. (b) Variation of charge current with dephasing strength for PC and APC. }
    \label{dep_MTJ}
\end{figure}

First, we simulate the dephasing effect on the trilayer MTJ device. We note that the TMR reduces as the dephasing strength is increased as shown in Fig. \ref{dep_MTJ}(a). As the TMR value depends on the currents, we have also analyzed the charge current for the PC and the APC as shown in Fig. \ref{dep_MTJ}(b). From the charge
current plot it is clearly seen that with an increase in dephasing strength, the currents in both the PC and the APC increase. While analyzing the current profile we notice that although currents in both configurations increase with $D_0$, the current in APC increases at a faster rate than that of PC, thus decreasing the TMR. For $D_0$=0.1 $eV^2$, the change in current in the PC is increased by 12.46\% and in the APC current is increased by 98.14\% with respect to $D_0$=0 $eV^2$, whereas for $D_0$=0.5 $eV^2$, the current increases by 17.8\% and 122.65\% respectively in the PC and the APC. From, \ref{TMR_eq} it is quite clear that due to the different rates of increase of current in PC and APC, the TMR reduces. So we get a trend that with an increase of dephasing strength, although the current increases, the TMR reduces in the trilayer case. The results obtained from our simulations are summarized in Table \ref{tab1}.
\begin{table}[htbp]
    \caption{Effects of dephasing on MTJ at 25 $nm^2$ cross-sectional area for V=0.01 V}
    \begin{center}
        \begin{tabular}{|c|c|c|c|}
            \hline
            $D_0(eV^2)$ & $I_p$ (mA) & $I_{AP}$ (mA) & TMR (\%)\\
            \hline
            0 & 0.1011$\times$ $10^{-3}$ & 0.1130$\times$ $10^{-4}$ &793.35\\
            \hline
            0.03 & 0.1083$\times$ $10^{-3}$ & 0.1763$\times$ $10^{-4}$ &514.62\\
            \hline
            0.1 & 0.1137$\times$ $10^{-3}$ & 0.2239$\times$ $10^{-4}$ &407.64\\
            \hline
            0.5 & 0.1191$\times$ $10^{-3}$ & 0.2516$\times$ $10^{-4}$ &373.22\\ 
            \hline
        \end{tabular}
        \label{tab1}
    \end{center}
\end{table}

\subsection{Pentalayer RTMTJ}
\begin{figure}[h]
    \subfloat[]{\includegraphics[scale=0.29]{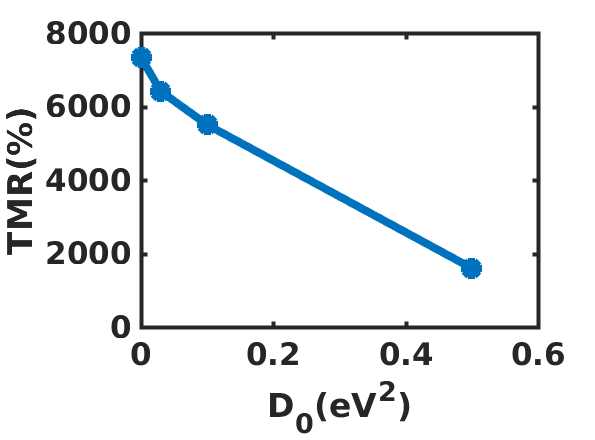}}
    \hfill
    \subfloat[]{\includegraphics[scale=0.28]{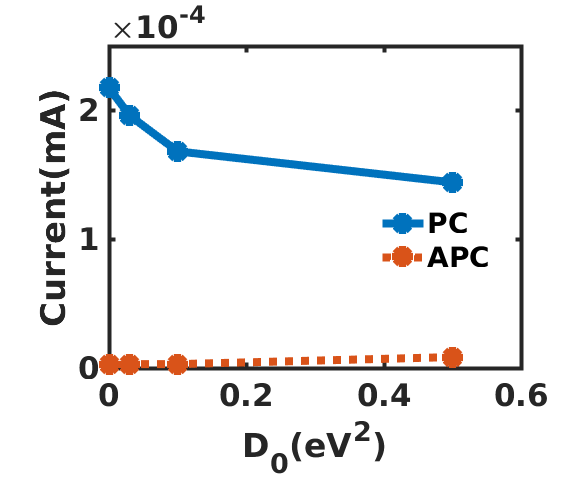}}
    \caption{Dephsing effects on RTMTJ. (a) Variation of TMR with strength of dephasing. (b) Variation of charge current with dephasing strength for PC and APC. }
    \label{dep_RTMTJ}
\end{figure}

\begin{figure}[h]
    \subfloat[]{\includegraphics[scale=0.21]{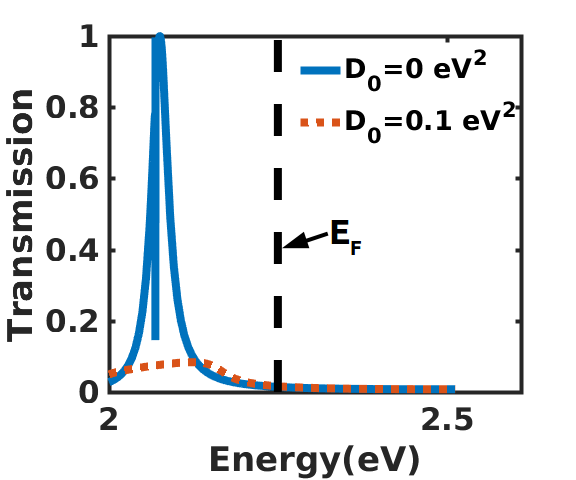}}
    \hfill
    \subfloat[]{\includegraphics[scale=0.21]{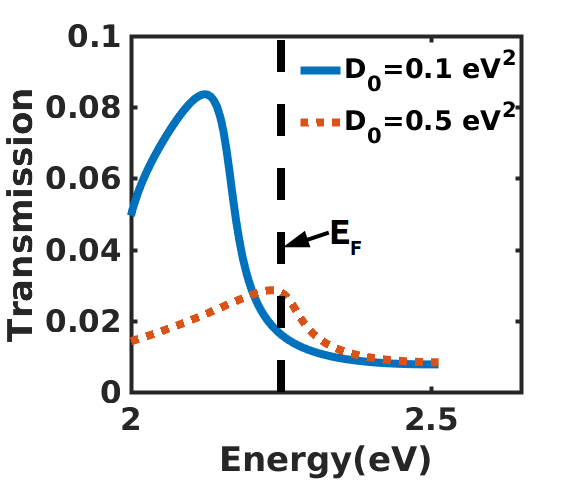}}
    \caption{Transmission spectrum for up-spin electrons for RTMTJ in PC for the lowest transverse mode. (a) Transmission spectrum for $D_0$=0 $eV^2$ and $D_0$=0.1 $eV^2$. (b) Transmission spectrum for $D_0$=0.1 $eV^2$ and $D_0$=0.5 $eV^2$. Vertical dashed line in both figures denote the Fermi energy level ($E_F$).}
    \label{trans_RTMTJ}
\end{figure}

Similarly, we investigate the effect of dephasing in pentalayer RTMTJs. Here we observe that the TMR reduces with an increase in dephasing strength in an almost linear fashion as shown in Fig. \ref{dep_RTMTJ} (a). Unlike the trilayer case, the reduction in TMR with dephasing is quite clear from the charge current variation with dephasing  
strength as shown in Fig. \ref{dep_RTMTJ} (b). From this figure, we can see that although the charge current in the APC increases with an increase of dephasing, the current in PC reduces drastically, which is why there is a sharp fall in the TMR with dephasing. For $D_0$=0.1 $eV^2$, the charge current in APC is increased by 2.54\%, but in PC, it is decreased by 22.68\% with respect to $D_0$=0 $eV^2$. However, for $D_0$=0.5 $eV^2$ the charge current in PC is decreased by 33.73\%, but in APC the current is increased by 189.6\%. From these numbers and from Eq. \ref{TMR_eq} it is quite clear to understand the reason behind this sharp fall in TMR with dephasing strength. The above mentioned numerical results for RTMTJ simulation are tabulated in Table \ref{tab2}.
\begin{table}[htbp]
    \caption{Effects of dephasing on RTMTJ at 25 $nm^2$ cross-sectional area for V=0.01 V}
    \begin{center}
        \begin{tabular}{|c|c|c|c|}
            \hline
            $D_0(eV^2)$ & $I_p$ (mA) & $I_{AP}$ (mA) & TMR (\%)\\
            \hline
            0 & 0.2175$\times$ $10^{-3}$ & 0.2927$\times$ $10^{-5}$ &7331.8\\
            \hline
            0.03 & 0.1960$\times$ $10^{-3}$ & 0.3011$\times$ $10^{-5}$ &6410.1\\
            \hline
            0.1 & 0.1682$\times$ $10^{-3}$ & 0.3001$\times$ $10^{-5}$ &5503.7\\
            \hline
            0.5 & 0.1441$\times$ $10^{-3}$ & 0.8476$\times$ $10^{-5}$ &1600.6\\ 
            \hline
        \end{tabular}
        \label{tab2}
    \end{center}
\end{table}
We know that the RTMTJ shows a very sharp transmission peak \cite{sharma2016ultrasensitive} for one type of spin near the Fermi level which gives rise to an ultrahigh TMR. From the transmission spectrum, we can get an idea about the currents in the devices. In order to find the reason behind this reduction of current with $D_0$, we investigate how the transmission spectrum changes with dephasing. The change in transmission spectra with dephasing strength for the up spin electron in PC for the lowest transverse mode are shown in Fig. \ref{trans_RTMTJ} (a) and (b). 
These plots reveal that the magnitude of the transmission peak reduces with the increase of dephasing strength and also the transmission spectrum get flattened, which matches with the previously reported trend \cite{sharma2018role}, and can hence be deduced as the reason behind the reduction of the charge current in PC with the increase of dephasing strength. 
\section{Conclusion}
We investigated how elastic dephasing effects influence the device characteristics for the trilayer MTJ and the pentalayer RTMTJ at the small transverse cross-sectional area. In the trilayer MTJ as well as the pentalayer RTMTJ, we note a fall in TMR with an increase in dephasing strength, where the reduction of TMR is much more drastic in the RTMTJ structure.  We attributed these trends by analyzing the variation of the transmission spectrum in order to intuitively explain this dramatic reduction. Although dephasing affects the TMR values for both devices, we note that the obtained TMR values are still in a reasonable range that may not hinder their usability for practical applications.


\begin{thebibliography}{1}

\bibitem{huai2008spin}
Y.~Huai, ``{Spin-Transfer Torque MRAM (STT-MRAM): Challenges and prospects},''
\emph{AAPPS bulletin}, vol.~18, no.~6, pp. 33--40, 2008.

\bibitem{Houssameddine2007}
D.~Houssameddine, U.~Ebels, B.~Dela{\"{e}}t, B.~Rodmacq, I.~Firastrau,
F.~Ponthenier, M.~Brunet, C.~Thirion, J.-P. Michel, L.~Prejbeanu-Buda, M.-C.
Cyrille, O.~Redon, and B.~Dieny, ``{Spin-torque oscillator using a
    perpendicular polarizer and a planar free layer.}'' \emph{Nature materials},
vol.~6, no.~6, pp. 441--447, 2007.

\bibitem{Sharma2015}
S.~Sharma, B.~Muralidharan, and A.~Tulapurkar, ``{Proposal for a domain wall
    nano-oscillator driven by non-uniform spin currents},'' \emph{Scientific
    Reports}, vol.~5, p. 14647, 2015.

\bibitem{Kiselev2003}
S.~I. Kiselev, J.~C. Sankey, I.~N. Krivorotov, N.~C. Emley, R.~J. Schoelkopf,
R.~a. Buhrman, and D.~C. Ralph, ``{Microwave oscillations of a nanomagnet
    driven by a spin-polarized current},'' \emph{Nature}, vol. 425, no. 6956, pp.
380--383, 2003.

\bibitem{Nikonov2011}
D.~E. Nikonov, G.~I. Bourianoff, and T.~Ghani, ``{Proposal of a spin torque
    majority gate logic},'' \emph{IEEE Electron Device Letters}, vol.~32, no.~8,
pp. 1128--1130, 2011.

\bibitem{parkin2004giant}
S.~S. Parkin, C.~Kaiser, A.~Panchula, P.~M. Rice, B.~Hughes, M.~Samant, and
S.-H. Yang, ``{Giant tunnelling magnetoresistance at room temperature with
    MgO (100) tunnel barriers},'' \emph{Nature materials}, vol.~3, no.~12, p.
862, 2004.

\bibitem{wang2010coherent}
W.~Wang, E.~Liu, M.~Kodzuka, H.~Sukegawa, M.~Wojcik, E.~Jedryka, G.~Wu,
K.~Inomata, S.~Mitani, and K.~Hono, ``{Coherent tunneling and giant tunneling
    magnetoresistance in $Co_2FeAl/MgO/CoFe$ magnetic tunneling junctions},''
\emph{Physical Review B}, vol.~81, no.~14, p. 140402, 2010.

\bibitem{chatterji2014enhancement}
N.~Chatterji, A.~A. Tulapurkar, and B.~Muralidharan, ``{Enhancement of
    Spin-transfer torque switching via resonant tunneling},'' \emph{Applied
    Physics Letters}, vol. 105, no.~23, p. 232410, 2014.

\bibitem{sharma2016ultrasensitive}
A.~Sharma, A.~Tulapurkar, and B.~Muralidharan, ``{Ultrasensitive nanoscale
    magnetic-field sensors based on resonant spin filtering},'' \emph{IEEE
    Transactions on Electron Devices}, vol.~63, no.~11, pp. 4527--4534, 2016.

\bibitem{mangin2006current}
S.~Mangin, D.~Ravelosona, J.~Katine, M.~Carey, B.~Terris, and E.~E. Fullerton,
``{Current-induced magnetization reversal in nanopillars with perpendicular
    anisotropy},'' \emph{Nature materials}, vol.~5, no.~3, p. 210, 2006.

\bibitem{meng2006spin}
H.~Meng and J.-P. Wang, ``{Spin transfer in nanomagnetic devices with
    perpendicular anisotropy},'' \emph{Applied physics letters}, vol.~88, no.~17,
p. 172506, 2006.

\bibitem{das2018scaling}
D.~Das, A.~Tulapurkar, and B.~Muralidharan, ``{Scaling Projections on
    Spin-Transfer Torque Magnetic Tunnel Junctions},'' \emph{IEEE Transactions on
    Electron Devices}, vol.~65, no.~2, pp. 724--732, 2018.

\bibitem{datta2}
S.~Datta, \emph{Quantum transport: Atom to Transistor}.\hskip 1em plus 0.5em
minus 0.4em\relax Cambridge University Press, 2005.

\bibitem{golizadeh2007nonequilibrium}
R.~Golizadeh-Mojarad and S.~Datta, ``Nonequilibrium Green's function based
models for dephasing in quantum transport,'' \emph{Physical Review B},
vol.~75, no.~8, p. 081301, 2007.

\bibitem{yanik2007quantum}
A.~A. Yanik, G.~Klimeck, and S.~Datta, ``{Quantum transport with spin
    dephasing: A nonequlibrium Green’s function approach},'' \emph{Physical
    Review B}, vol.~76, no.~4, p. 045213, 2007.


\bibitem{Datta2012}
D.~Datta, B.~Behin-Aein, S.~Datta, and S.~Salahuddin, ``{Voltage asymmetry of spin-transfer torques},'' \emph{IEEE Transactions on Nanotechnology},
vol.~11, no.~2, pp. 261--272, 2012.


\bibitem{sharma2018role}
A.~Sharma, A.~A. Tulapurkar, and B.~Muralidharan, ``Role of phase breaking
processes on resonant spin transfer torque nano-oscillators,'' \emph{AIP
    Advances}, vol.~8, no.~5, p. 055913, 2018.


\end{thebibliography}
\end{document}